# SWELLING NEGATION DURING SINTERING OF STERLING SILVER: AN EXPERIMENTAL AND THEORETICAL APPROACH

Charles Manière[a]*, Elia Saccardo[b], Geuntak Lee[a, c], Joanna McKittrick[c], Alberto Molinari[b], Eugene A. Olevsky[a, d]

(a) Powder Technology Laboratory, San Diego State University, San Diego, USA
(b) Department of Industrial Engineering, University of Trento, Trento, Italy
(c) Mechanical and Aerospace Engineering, University of California San Diego, La Jolla, USA
(d) NanoEngineering, University of California, San Diego, La Jolla, USA

**Keywords**

Silver; Sintering; Swelling; Modeling, Dilatometry; Atmosphere control

**Abstract**

One of the main challenges of the sintering of sterling silver is the phenomenon of swelling causing de-densification and a considerable reduction of the sintering kinetics. This swelling phenomenon opposes sintering and it needs to be addressed by a well-controlled processing atmosphere. In the present study, the pressure-less sintering behavior of sterling silver is investigated in air, argon, and vacuum. A specially modified spark plasma sintering mold is designed to study the pressure-less sintering of sterling silver in a high vacuum environment. The conducted analysis is extended to the new constitutive equations of sintering enabling the prediction of the swelling phenomena and the identification of the internal equivalent pressure that opposes the sintering.

* Corresponding author: CM: Powder Technology Laboratory, San Diego State University, 5500 Campanile Drive, San Diego, CA 92182-1323,
E-mail address: cmaniere@mail.sdsu.edu

# Graphical Abstract

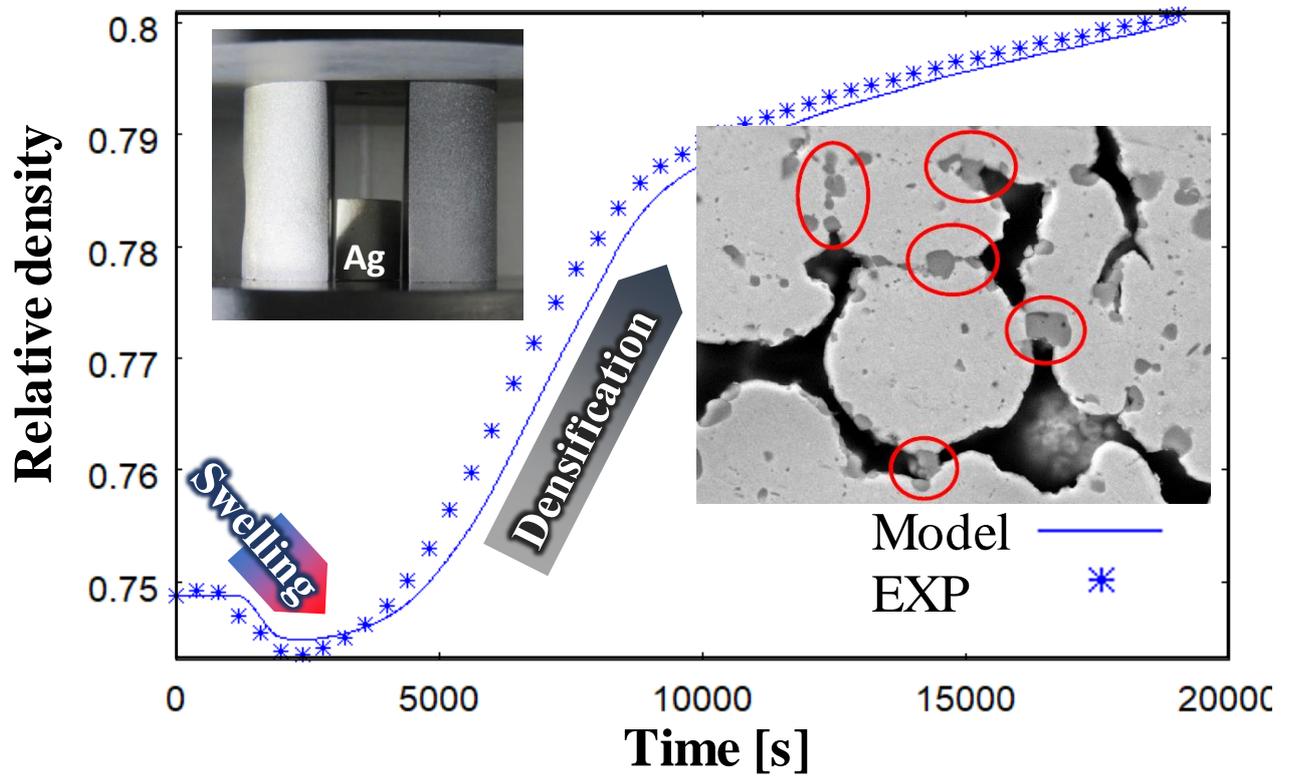

# Highlights:

- Pressure-less sintering of silver under various atmospheres.
- Overcoming swelling during high vacuum sintering.
- Modelling the competition between densification and swelling.

## 1. Introduction

Sterling silver is an alloy of silver containing 7.5 weight % of other metals (mainly copper) to reinforce its mechanical properties. This alloy is used mostly in jewelry, thanks to its shining aspect. The resistance of silver to corrosion makes this metal also interesting for low and elevated temperature electronics, electrical contact parts, high quality mirrors, etc. Numerous investigations are dedicated to the influence of the additives on the mechanical properties, tarnish and corrosion resistance of sterling silver [1–3].

The sintering of sterling silver is also highly influenced by the presence of the additives like copper. Phenomena of oxidation and swelling causing de-densification and abnormal porosity evolution have been reported for the sintering of silver and copper powders [4–6]. Swelling is often associated with the presence of a trapped gas inside the pores generated by the desorption of impurities such as oxygen, carbon, nitrogen or hydrogen during heating [4,7–9]. This gas pressure creates a force that opposes the sintering stress resulting from capillarity forces during sintering. This competition between the sintering forces and swelling is a complex phenomenon where the role of the initial relative density, green sample preparation, heating regime and the impurities is crucial [4,8]. Olmos *et al* [5] have characterized the swelling of a copper powder specimen by in situ X-ray microtomography and showed the complexity of the phenomenon generating interconnected porosity by micro-channels and a large distribution of pore sizes. Particularly, they highlighted the impact of the initial powder particle distribution and packing that can initiate abnormal porosity growth. Lin *et al* [6] have reported also the presence of in situ generated cracks in copper particles. From the point of view of the mass transport mechanism of sintering with swelling, it is expected that the grain boundary diffusion is still the dominant mechanism for pressure-less sintering [10]. However, the kinematic of this mechanism may be limited by the opposing sintering pressure or the eventual disturbance of the surface diffusion due to the impurities and the change in the porosity skeleton [11]. The impurities may also generate nano size precipitates which can potentially disturb the sintering kinetics [12].

The sintering of silver in contrast with classical casting methods is actually of an increasing interest considering the appearance of technologies able to generate highly complex shape green samples, like selective laser sintering / melting – assisted 3D printing, binder jetting and other additive manufacturing techniques [13–15]. The control of the swelling is then crucial to obtain dense specimens by sintering. In this study, the first part is dedicated to the influence of the atmosphere (air, argon and vacuum) on the sintering of sterling silver. The evidence of the swelling phenomenon and the way to avoid it is investigated. The second theoretical part of the paper is focused on the quantification and modeling of this swelling phenomenon. In particular, the so called "equivalent opposing sintering pressure" is determined. The continuum theory of sintering [16,17] is employed and all the sintering/swelling parameters are determined using the experimental data obtained for different atmospheres to ensure the comprehensive coupled modeling of both interconnected phenomena of sintering and swelling.

## 2. Experiments

The sterling silver powder composition is 92.5 % (wt.) of silver and 7.5 % (wt.) of copper. At this composition, the silver alloy has a solidus temperature of 788 °C ($\beta$ solid solution $\rightarrow$ $\beta$ + liquid) and a liquidus temperature of 894 °C ($\beta$ + liquid $\rightarrow$ liquid) [18]. To avoid any liquid phase during sintering, the specimen temperatures have been limited to 770 °C. The sterling silver was supplied by Progold S.p.A. (Italy) and includes particles with diameter ranging from 5 to 30 μm. The mean radius was calculated through the scanning electron microscopy (SEM *FEI Quanta 450, USA*), the image of the loose powder is reported in Figure 1. The resulting mean diameter is 6 ± 4 μm. The powder has been made using the gas atomization technique, therefore the powder particles are well rounded.

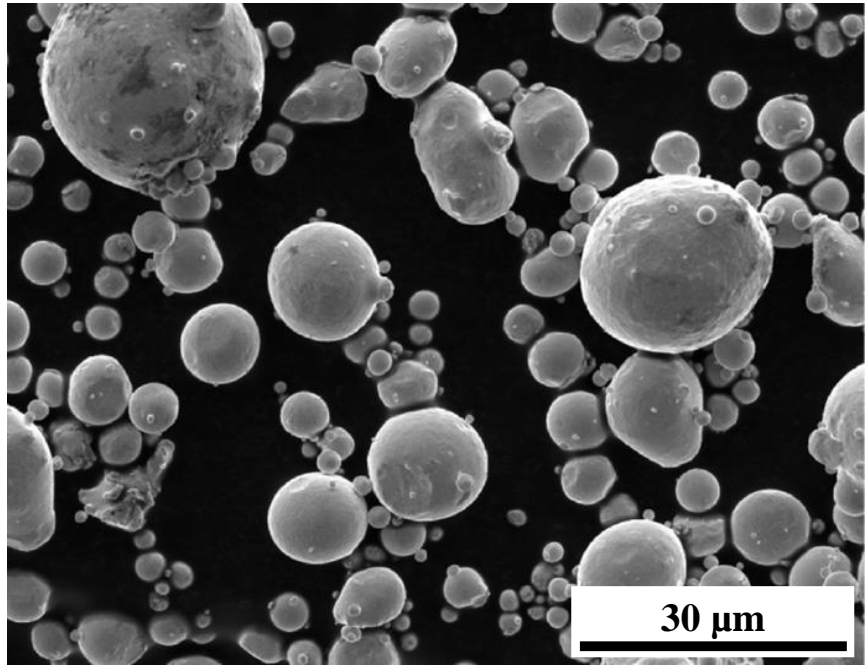

*Figure 1. SEM image of sterling silver powder.*

The sintering of sterling silver is investigated under the three different atmospheres, air, argon, and vacuum. The green samples were prepared by a Spark Plasma Sintering (*SPSS DR.SINTER Fuji Electronics model 515*) ramping tests (25 k/min, 50 MPa) interrupted at 350 °C. This approach allows obtaining 73-75 % dense green samples without a binder. The sintering in air and argon has been carried out in the dilatometer (*Unitherm model 1161, Anter Corporation)*. The sintering shrinkage curves have been calculated using the displacement curves obtained by an alumina pushrod in contact with the sample. The thermal expansion of the tooling has been subtracted from the sintering displacement by another alumina pushrod placed on the support next to the sample. The specimen temperature was measured by a S-type thermocouple placed next to the sample. For the experiment under vacuum, a modified spark plasma sintering pressure-less configuration has been specifically adapted to allow the heating and sintering of the sample under conditions of vacuum (20 Pa). The heating is generated by the two graphite heating elements placed on both sides of the silver sample (Figure 2a). The temperature of the sample is PID regulated by a K-type thermocouple located at the basis of the sample (Figure 2b) where a 2 mm deep hole is made to allow a direct contact of the sample and the thermocouple without any disturbance of the sample densification. The heating elements have been placed with an aperture in the center to measure the evolution of the sample geometry by a camera located outside the vacuum chamber. A hole was created in one of the graphite heating elements to allow a lateral illumination of the sample in order to amplify the contrast of the sample and the background of the images at low temperatures (Figure 2c). At high temperatures (from 570 °C to 770 °C), the sample emits a red-orange light and the geometrical dimensions can be determined by the direct contrast with the background which is darker. The average sample dimensions have been determined by the three measurements of the diameter and height of the images recorded during the test (Figure 2d). The discrepancies between the dilatometer sintering kinetics and the pressure-less SPS kinetics are mainly originated from the displacement curves where the error is of the order of 1 µm for the dilatometer and 10 µm for the SPS setup.

For the three experiments, the same thermal cycle has been imposed on the sample. An initial and fast heating of 25 K/min is imposed up to 550 °C, no sintering occurs in this range of temperatures, then a heating ramp of 2 K/min is imposed up to a temperature of 770°C (below the silver alloy solidus). The SPS experiment (in vacuum) has a time limit that corresponds roughly to the end of the 2 K/min ramp at 770 °C. For the sintering in air and argon performed with the dilatometer, three hours of holding at 770 °C are added.

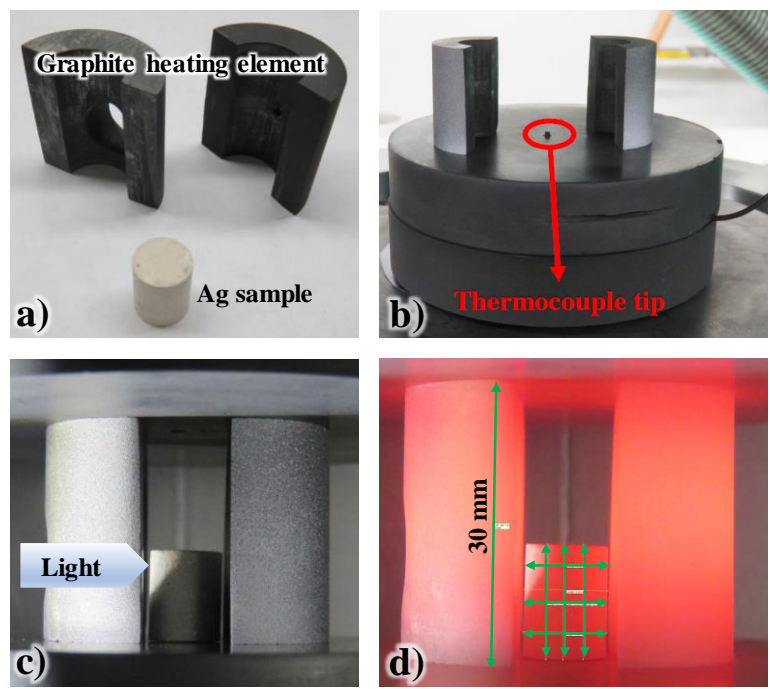

*Figure 2. Pressure-less vacuum SPS configuration with: (a) the Ag sample and graphite heating elements, (b) the thermocouple location, and the assembly at (c) low and (d) high temperature.*

## 3. Results and discussion

The results of the sintering of silver in air, argon and vacuum will be presented first. In the second part, the sintering and the swelling model will be detailed. The identification of the sintering parameters will be investigated first by the regression method and by comparison with the master sintering curve approach. Then, the determination of the equivalent internal "opposing sintering pressure" responsible for swelling will be carried out.

*3.1. Sintering of silver under different atmospheres*

The sample recorded temperature cycle and the relative density curves for the experiments in air, argon and vacuum (pressure-less SPS configuration) are reported in Figure 3. The experiments in air and argon present similar densification curves. At the first stage, between 400 °C and 600 °C, both samples exhibit swelling up to 2 % of de-densification. After 600 °C, the densification of the samples starts. The argon sample presents a faster sintering rate than the one of the air sample under the 2 K/min ramp; during the 3 hours of holding this difference is decreased.

The sample sintered in vacuum using the modified pressure-less SPS configuration presents a very different sintering behavior. The data points for this vacuum experiment, which are determined by successive optical images, have some obvious level of scatter (few percent), however, the sintering trend is clearly very fast compared to the air and argon experiments. The swelling phenomenon seems inexistent. At the end of the 2 K/min ramp, the vacuum sintered sample's densification reaches the substantial value of 92 %. This represents three times the amount of the densification of the air and argon sintered samples which reached, respectively, 79 % and 80 % of the final densification after three more hours of holding at 770 °C.

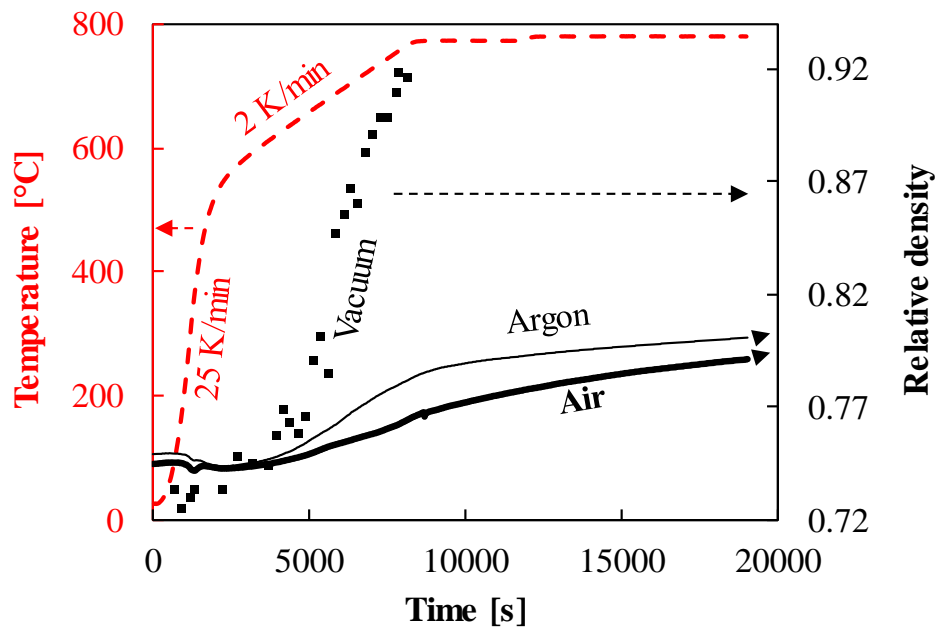

*Figure 3. Relative density curves (in black) of silver samples in vacuum, argon and air, the same thermal cycle is imposed onto the samples and is reported in red.*

The microstructures of the three sintered samples in the center and the edge are reported in Figure 4. The sample sintered in air has a very inhomogeneous microstructure with a central area well sintered and a thick porous external shell (2.5 mm on the 9 mm diameter sample). The sample sintered in argon has a homogeneous but porous microstructure (about 20 % of porosity based on the sample geometrical density). Finally, the sample sintered in vacuum has a homogeneous and well-sintered microstructure compared to the other tests. The sintering under vacuum is clearly the way to sinter the sterling silver powder.

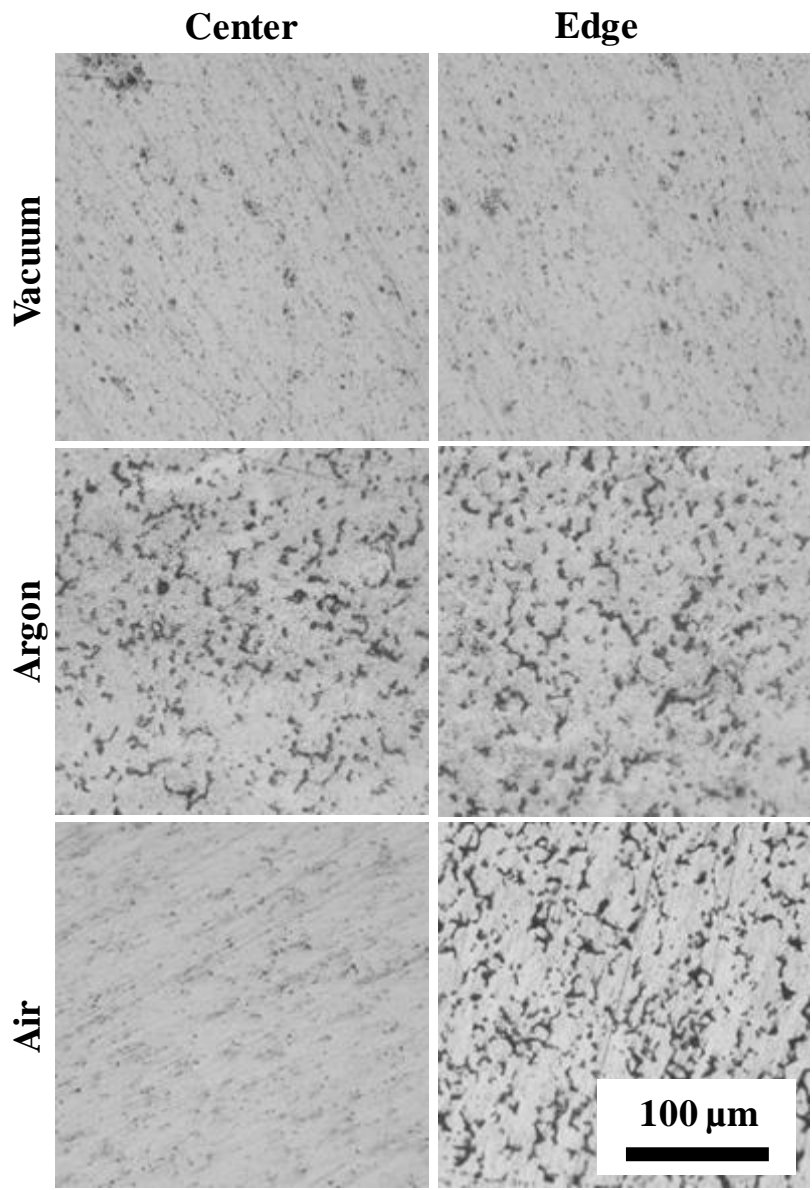

*Figure 4. Microstructures of the sintered silver samples in vacuum, argon and air in the center and the edge of the specimens (with black areas indicating the porosity), thermal cycle: 25 K/min to 550°C, 2 K/min to 770°C and 3 hours of holding at 770°C.*

There are two main phenomena that can explain the low densification in air and argon.

The first is the swelling that is observed in both samples at temperatures between 400 °C and 600 °C; this phenomenon can be explained by the presence of a trapped gas originated from the processing atmosphere or the presence of impurities in the closed pores (closed intergranular porosity network, intragrain porosity and intraparticle porosity also called secondary porosity) [4,8]. The presence of at least a small fraction of closed porosity in the beginning of the sintering should contribute to the possible origin of swelling due to the gas pressure. Because there is no significant grain growth and the elimination of intragrain pores is very difficult by pressure-less sintering, it is possible to estimate the fraction of the closed intragrain porosity from the final sintered microstructure [19]. Based on the microstructure (figure 4) obtained in Argon atmosphere, which contains both inter- and intra-grain porosity; the intergranular porosity is about 20 % and the intragrain porosity is 1-5 %. There are then potentially few percent of closed porosity (initially present or in situ generated) or more since the small size pores are difficult to observe.

The second phenomenon is the oxidation that can affect the copper contents of the alloy and then create an oxide phase that can form rigid inclusions which increase the equivalent material viscosity and then slow down the sintering [20]. The curves in Figure 3 and microstructures in Figure 4 suggest that the significant level of vacuum of the SPS machine (20 Pa) prevents the powder oxidation and facilitates the elimination of the internal gas (trapped or produced by the impurity desorption). Then, the argon atmosphere prevents the sample oxidation but the swelling is still present, and the competition of the internal opposing sintering pressure (possibly explained by trapped gas) and sintering stress slowdowns the densification. The oxidation seems to be the factor responsible for inhomogeneous microstructure of the sample sintered in air. The Energy-dispersive X-ray spectroscopy analysis (EDX) of the air and vacuum sintered samples is comparatively represented in Table 1 for the sample surface and center. The comparison shows that the edge of the air sintered sample has less silver content and more oxygen compared to the center. On the contrary, the composition of the vacuum specimen is more homogeneous. The porous shell of the air sintered sample has been analyzed by SEM and EDX in more detail (Figure 5). The microstructure and EDX analyses show that the copper phase of the sterling silver alloy is almost completely segregated and transformed into copper oxide that appears at the particles' surfaces and grain boundaries. The other (white color) phase is mainly composed of silver. This confirms the presence of the copper oxide inclusions. These inclusions, like in most oxide dispersed strengthened alloys (ODS), reinforce the material resistance to deformation and then to sintering [21]. This can explain the inhomogeneous microstructure, where the surface of the sample is oxidized, strengthened and subjected to swelling. In contrast, the central part is more protected from the oxidation and can be sintered more easily.

Table. 1: Energy-dispersive X-ray spectroscopy of vacuum and air sintered samples in the center and the surface of the specimen.

| EDX | Vacuum | | Air | |
|---|---|---|---|---|
| Atomic % | Center | Surface | Center | Surface |
| O | 14.4 | 32.3 | 29.1 | 59.6 |
| Cu | 9.94 | 14.8 | 7.97 | 22.6 |
| Ag | 75.7 | 53.6 | 62.9 | 18.1 |

The sintering of silver in air is therefore prohibited but the argon case is more interesting for the understanding of the sintering process. In this case the atmosphere oxygen content is poor, and the sintering inhomogeneity that appears in air is not present. However, the sintering is anyway drastically reduced compared to the vacuum case. Moreover, the swelling is the most intense compared to all the specimens. In this case the competition between the internal gas pressure and the sintering stress is a crucial mechanism, sufficiently strong to prevent the complete densification of the sample despite the absence of oxygen. The determination of the internal opposing sintering pressure responsible for the swelling and the modeling of the sintering behavior are discussed in the next sections.

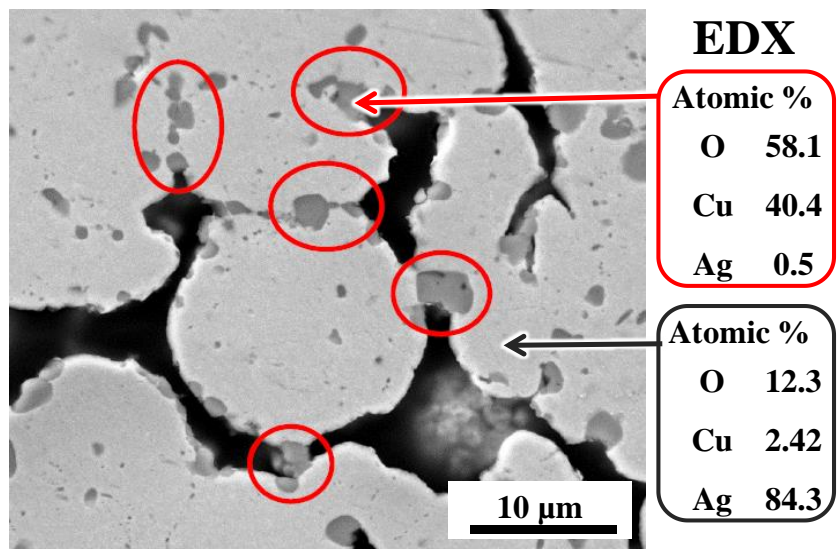

*Figure 5. Edge microstructure and EDX analysis of the air sintered silver sample.*

*3.2. Sintering model without swelling*

The pressure-less sintering can be defined by the continuum theory of sintering [16] and the following equation:

$$-P_l = \frac{A\psi\dot{\theta}}{(1-\theta)} \quad (1)$$

where $P_l$ is the sintering stress (Pa), $A$ the material viscosity (Pa s), $\psi$ the bulk modulus, and $\theta$ the porosity.

The expressions of $P_l$, $\psi$ can be defined theoretically [22] as:

$$P_l = \frac{3\alpha(1-\theta)^2}{r_0} \quad (2)$$

$$\psi = \frac{2(1-\theta)^3}{3\theta} \quad (3)$$

where $\alpha$ is the surface energy (J m$^{-2}$) and $r_0$ is the particle radius (m).

The temperature dependence of the material viscosity can be defined as [16,23]:

$$A = A_0 T \exp\left(\frac{Q}{RT}\right) \quad (4)$$

where $T$ is the temperature ($K$), $A_0$ is a preexponential constant (Pa s K$^{-1}$), $R$ is the gas constant (J K$^{-1}$ mol$^{-1}$) and $Q$ is the activation energy (J mol$^{-1}$).

Inserting equations (2), (3) and (4) in (1) we obtain the analytical expression describing the case of pressure-less sintering (5), where the three materials unknown parameters to be determined are $A_0$, $\alpha$ and $Q$:

$$\frac{-3(1-\theta)^3}{r_0 T \psi \dot{\theta}} = \frac{A_0}{\alpha} \exp\left(\frac{Q}{RT}\right) \quad (5).$$

Our strategy is to identify the sintering parameters in a processing regime, which is not influenced by the swelling, using the vacuum experiment sintering data of Figure 3. We can then determine the ratio $A_0/\alpha$ and the activation energy $Q$ by the linear regression equation:

$$\ln\left(\frac{-3(1-\theta)^3}{r_0 T \psi \dot{\theta}}\right) = \ln\left(\frac{A_0}{\alpha}\right) + \frac{Q}{RT} \quad (6).$$

The left-side term can be calculated using $r_0 = 3E-6$ m; the porosity rate ($\dot{\theta}$) is determined from the data in Figure 3 via a polynomial smoothing function. The regression graph is reported in Figure 6. This regression has been done using the curve (see figure 3) corresponding to vacuum processing in the 2 K/min ramp between 0.75 and 0.87 of relative density range.

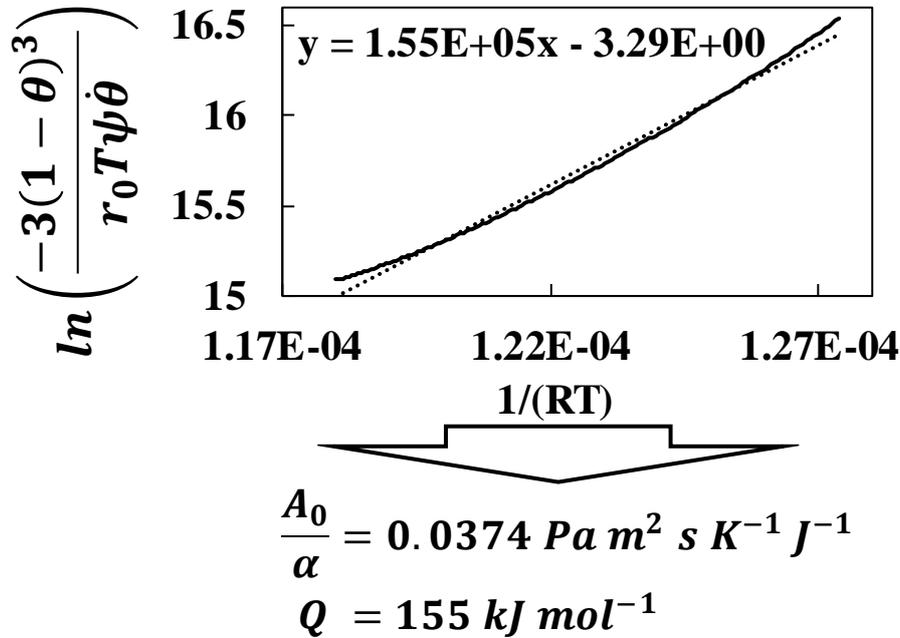

Figure 6. Determination of the pressure-less sintering parameters by regression (the dotted curve is the fitting curve).

Another experiment has been done using the SPS pressure-less configuration at 5 K/min. The experimental data of the sintering curves at 2 and 5 K/min (reported in Figure 7a) can be used together to determine another independent sintering activation energy via the master sintering curve method (MSC) [24,25]. The activation energy obtained by regression and MSC method can then be compared. The MSC method requires plotting relative density using the expression:

$$ln(\vartheta(t,T)) = ln\left(\int_0^t \frac{1}{T} exp\left(\frac{-Q}{RT}\right) dt\right) \qquad (7).$$

Both 2 and 5 K/min curves should then match together for the properly chosen value of the material activation energy. A minimization method is commonly employed to determine the correct activation energy by minimizing the fitting error of the curves. The MSC method' results are reported in Figure 7b. The activation energy is about 177 kJ/mol, which is close to the value of 155 kJ/mol identified by the regression. Knowing the dispersion of the experimental data points (Figure 7a), this result appears to be quite reasonable.

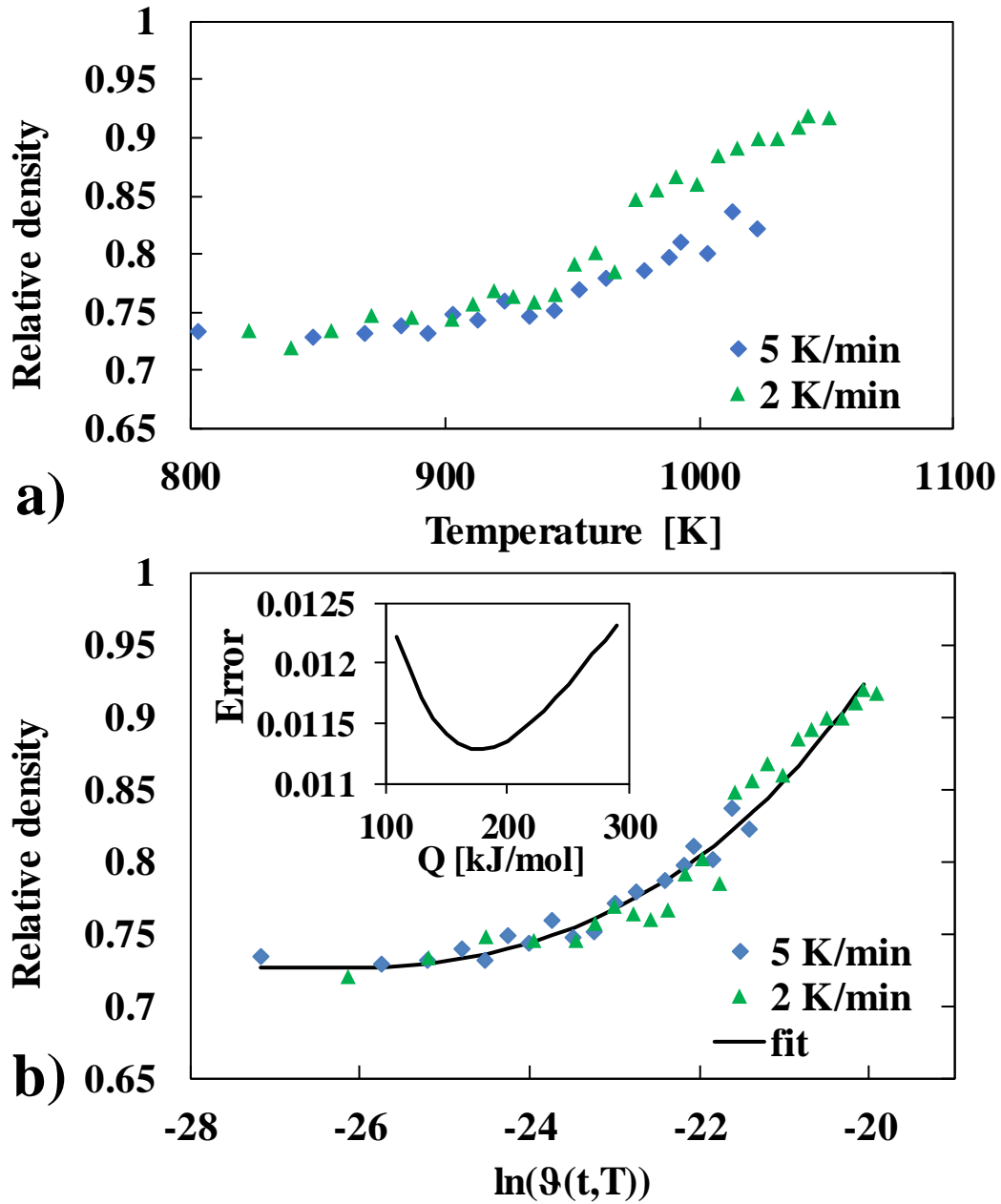

*Figure 7. (a) Vacuum sintered densification curves at 2 and 5 K/min, (b) determination of the sintering activation energy by master sintering curve approach.*

From the deformation mechanism maps data [26], the activation energy of silver is 185 kJ/mol for lattice diffusion and 90 kJ/mol for grain boundary diffusion. The values of activation energy for sterling silver (155-177 KJ/mol) seem to indicate the lattice diffusion mechanism. The tendency of the copper phase to segregate and/or oxidize (Figure 5) at the grain boundaries can be an explanation for this mechanism. Indeed, copper is used to strengthen the silver and copper oxide can form rigid inclusions. The segregation of these components at the grain boundaries can disturb the grain boundary diffusion and act in favor of the lattice diffusion mechanism through the silver grains' volume.

*3.3. Sintering model with swelling*

So far, the sintering behavior of the sterling silver powder is determined without swelling. The modeling can be extended to take into account the swelling phenomenon. For these investigations, the data of sintering in argon (Figure 3) will be considered; the sintering in air represents an even more complex problem that should also include the phenomenon of oxidation and is not explored in this study.

The sintering model with swelling includes an additional term $Ps$ in the constitutive equation of sintering (1), which is the equivalent internal pressure that opposes the sintering stress $P_l$. This term ($Ps$) has an opposite sign of the sintering stress ($P_l$) because both phenomena compete with each other. The swelling happens only if $Ps>P_l$. The new sintering constitutive equation taking into account the swelling phenomenon is:

$$P_s - P_l = \frac{A\psi\dot{\theta}}{(1-\theta)} \qquad (8).$$

The term $Ps$ may evolve with time and sample temperature. Similarly to the determination of $A_0$, the determination of the internal opposing sintering stress ($Ps$) should be done via the ratio $Ps/\alpha$ with the unknown surface energy $\alpha$. The equation used to determine experimentally this ratio $Ps/\alpha$, based on the already determined sintering constitutive parameters ($A_0/\alpha$ and $Q$), is:

$$\frac{P_s}{\alpha} = \frac{A_0}{\alpha}exp\left(\frac{Q}{RT}\right)\frac{T\psi\dot{\theta}}{(1-\theta)} + \frac{3(1-\theta)^2}{r_0} \qquad (9).$$

The ratio of the swelling gas pressure $Ps/\alpha$ is determined for the sintering in argon experiment, and the results are reported in Figure 8a. The value of the surface energy is unknown; however, this value is often reported to be close to unity [10,27] (*α~1*), the ratio *Ps/α* is therefore close to *Ps*. We can then estimate the values of the opposing sintering pressure to be between 0.5 and 2.5 MPa. During heating, the peak of opposing sintering pressure up to 2.5 MPa is observed between 200 and 600 °C, and this explains the swelling phenomenon that occurs for this range of temperatures. During the temperature holding, an opposing sintering pressure of about 0.6 MPa is needed to explain the sintering slowdown compared to the vacuum sintering experiment. The final model of sintering and swelling is reported in Figure 8b. The model predicts a reasonable estimation for both the swelling and sintering phenomena.

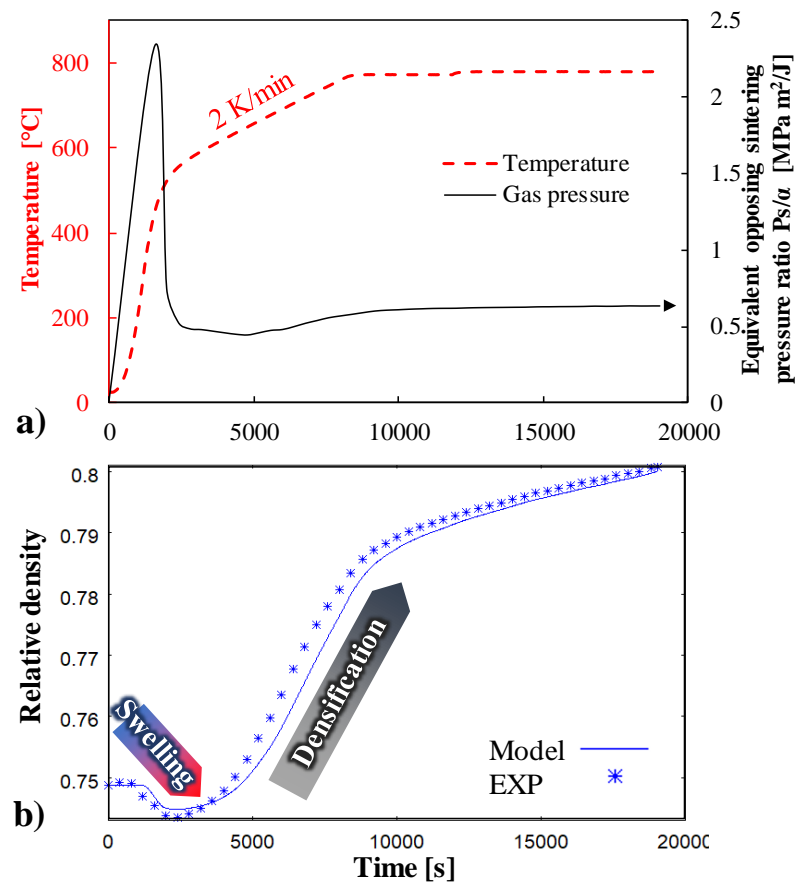

*Figure 8. (a) Sample temperature and equivalent opposing sintering pressure ratio Ps/α for the argon sintered sample, (b) modeling results for the sintering in argon test including the internal gas pressure term versus experimental data.*

## 4. Conclusions

The pressure-less sintering of sterling silver suffers from a very restrictive phenomenon of swelling which is responsible for the slowdown of the sintering kinetics and the de-densification of the sample. The control and comprehension of this phenomenon have been investigated on the one hand, throughout the sintering in air, argon, vacuum and, on the other hand, by modeling. The conducted work renders the following three main conclusions:

- The sintering under air and argon exhibits phenomena of swelling and oxidation. The sintering in air is very inhomogeneous and renders a porous and oxidized shell and a dense core specimen. The sintering in argon is more homogeneous but the swelling phenomenon is more intense. Both air and argon sintered samples manifest significant reduction of the sintering kinetic compared to the sintering in vacuum case.

- The sintering under a significant level of vacuum (20 Pa) is considerably faster, presents no apparent swelling and provides homogeneous densification. The presence of copper-rich zones at the grain boundaries and the analysis of the densification data seem to indicate that lattice diffusion is a dominant sintering mechanism.

- The significant difference of the sintering kinetics between the air/argon and vacuum sintered samples indicates the probable presence of in-particle trapped gas that causes the swelling and slowdown of the sintering. Using a new constitutive model of sintering-swelling, we have estimated for the argon case that this equivalent opposing sintering pressure (gas pressure) is close to 2.5 MPa in the swelling regime and 0.6 MPa in the sintering regime.

The sintering of sterling silver without swelling is possible but high vacuum conditions appear to be required to ensure fast shrinkage and the removal of the residual impurities causing swelling and oxidation phenomena.

**Acknowledgments:**

The support of the US Department of Energy, Materials Sciences Division, under Award No. DE-SC0008581 is gratefully acknowledged.

# Table captions:

Table. 1: Energy-dispersive X-ray spectroscopy of vacuum and air sintered samples in the center and the surface of the specimen.

# Figure captions:

Figure 1. SEM image of sterling silver powder.

Figure 2. Pressure-less vacuum SPS configuration with: (a) the Ag sample and graphite heating elements, (b) the thermocouple location, and the assembly at (c) low and (d) high temperature.

Figure 3. Relative density curves (in black) of silver samples in vacuum, argon and air, the same thermal cycle is imposed onto the samples and is reported in red.

Figure 4. Microstructures of the sintered silver samples in vacuum, argon and air in the center and the edge of the specimens (with black areas indicating the porosity), thermal cycle: 25 K/min to 550°C, 2 K/min to 770°C and 3 hours of holding at 770°C.

Figure 5. Edge microstructure and EDX analysis of the air sintered silver sample.

Figure 6. Determination of the pressure-less sintering parameters by regression method (the dotted curve is the fitting curve).

Figure 7. Vacuum sintered densification curves at 2 and 5 K/min, (b) determination of the sintering activation energy by master sintering curve approach.

Figure 8. (a) Sample temperature and equivalent opposing sintering pressure ratio $P_s/\alpha$ for the argon sintered sample, (b) modeling results for the sintering in argon test including the internal gas pressure term versus experimental data.